\documentstyle[12pt]{article}
\newcommand{\be}{\begin{equation}}
\newcommand{\bea}{\begin{eqnarray}}
\newcommand{\eea}{\end{eqnarray}}
\newcommand{\ee}{\end{equation}}
\newcommand{\lb}{\label}
\begin{document}

\begin{titlepage}
\begin{flushright}
Freiburg THEP 94/11\\
gr-qc/940540
\end{flushright}
\vskip 1cm
\begin{center}
{\large\bf  WHEELER-DEWITT METRIC AND THE ATTRACTIVITY OF GRAVITY}
\vskip 1cm
{\bf Domenico Giulini and Claus Kiefer}
\vskip 0.4cm
 Fakult\"at f\"ur Physik, Universit\"at Freiburg,\\
  Hermann-Herder-Str. 3, D-79104 Freiburg, Germany.
\end{center}
\vskip 2cm
\begin{center}
{\bf Abstract}
\end{center}
\begin{quote}
We investigate the class of ultralocal metrics on the configuration
space of canonical gravity.
It is described by a parameter $\alpha$, where $\alpha=0.5$
corresponds to general relativity.
 For $\alpha$ less than a critical value
the signature is positive definite, while for all other values it is
indefinite. We show that in the positive definite case gravity
becomes repulsive. From the primordial
helium abundance we find that $\alpha$ must lie between $0.4$
and $0.55$.
 \end{quote}
\vskip 2cm
\begin{center}
{\em To appear in Phy. Lett. A}
\end{center}

\end{titlepage}

In observations of phenomena both within and outside the solar system,
the general theory of relativity has passed many viable tests in the
last decades. This success, together with the conceptual simplicity
and beauty of the theory, is the reason why it is commonly regarded
as the fundamental theory of gravity. And yet,
it is desirable to study alternative theories of gravity, for the following
reasons. First, by recovering general relativity from a more general,
alternative, theory in an appropriate limit one can {\em quantify}
its success by excluding parameter values referring to the alternative
theory. The perhaps best known example is the Jordan-Brans-Dicke theory
which in addition to the metric contains a scalar field playing the
role of a dynamical gravitational ``constant." The coupling of this
additional field to matter is specified by a parameter $\omega$
which must satisfy, from Viking radar measurements,
 the observational constraint
 $\omega >600$ ($\omega\to\infty$ recovers general
relativity). The second reason is quantum theory. Since all other known
interactions are successfully described in a quantum framework, one expects
that a quantum theory of gravity is also more fundamental than its
classical counterpart. Such a theory could then in principle lead
to classical limit different from general relativity. One example
is superstring theory with its prediction of dilaton fields
(similar to the Jordan-Brans-Dicke field) and antisymmetric tensor
fields in addition to the metric.

While most alternative theories, such as scalar-tensor theories or
higher derivative theories, are given in a Lagrangean prescription,
the viewpoint put forward here will be
the opposite one. The motivation is provided by canonical quantum
gravity in its geometrodynamical version, where the central kinematical
concept is a wave functional on superspace, the configuration space
of all three-geometries. By {\em varying} the metric on this configuration
space, we obtain a class of alternative theories even in the classical
limit (to which the discussion will be restricted). The study
of these alternative theories will provide an interesting new viewpoint
on general relativity itself.

The central role in canonical gravity is played by the Hamiltonian
constraint (with $c=1$)
\be H=\int d^3xN\sqrt{h}\left(\frac{16\pi G}{\sqrt{h}}
    G_{abcd}\pi^{ab}\pi^{cd}-\frac{R-2\Lambda}{16\pi G}
    +{\cal H}_m\right) =0, \lb{1} \ee
where $R$ is the Ricci scalar on a three-dimensional space,
$\Lambda$ is the cosmological constant, and $N$ is the lapse function
for which we will make the choice $N=1$ in the following. The coefficients
$G_{abcd}$ in front of the geometrodynamical momenta depend on the
three-metric $h_{ab}$ and play themselves the role of a metric on
$\mbox{Riem}\Sigma$, the space of all three-metric associated with a
manifold $\Sigma$. They are called {\em DeWitt metric} and are given
by the expression
\be G_{abcd}=\frac{1}{2\sqrt{h}}(h_{ac}h_{bd}+ h_{ad}h_{bc}
    -h_{ab}h_{cd}). \lb{2} \ee
We now {\em modify} this metric through the introduction of a
parameter $\alpha$,
\be G_{abcd}^{\alpha}=\frac{1}{2\sqrt{h}}(h_{ac}h_{bd}+ h_{ad}h_{bc}
    -2\alpha h_{ab}h_{cd}). \lb{3} \ee
We refer to the general class (3) of metrics as ``Wheeler-DeWitt metrics"
since the quantum version of the constraint (1),
$\hat{H}\Psi=0$, is referred to as the
Wheeler-DeWitt equation. The metrics (3) exhaust the class of all
ultralocal metrics on $\mbox{Riem}\Sigma$, i.e. metrics which do not
contain space derivatives \cite{Wi}.
 We note in bypassing that the corresponding
metric in the space of connections, which appears in the Ashtekar
approach to canonical gravity, does not possess this property of being
ultralocal.

The inverse metric to (3) is given by the expression
\be G^{abcd}_{\beta} =\frac{\sqrt{h}}{2}(h^{ac}h^{bd}
    +h^{ad}h^{bc}-2\beta h^{ab}h^{cd}), \lb{4} \ee
where
\be \alpha+\beta =3\alpha\beta. \lb{5} \ee
In the case of general relativity, which corresponds to the choice
$\beta=1$ ($\alpha=1/2$), the signature of (3) is, at each point
of space, given by $\mbox{diag}(-,+,+,+,+,+)$. Because of this indefinite
signature, vectors in $\mbox{Riem}\Sigma$ may be ``lightlike",
i.e. they may have vanishing scalar product with respect to
(4). Connected to
the existence of such lightlike vectors is the fact that
 one cannot project everywhere this metric down to a metric
on superspace, i.e. the space $\mbox{Riem}\Sigma$ modulo
diffeomorphisms \cite{Ni,Hi}. On the other hand, there are sets
in superspace where the infinitely many minus signs in DeWitt's metric
reduce to {\em one} global minus sign. This happens, for example,
for metrics which describe a closed Friedmann universe.
The Wheeler-DeWitt equation is then truly hyperbolic, a fact which has
important consequences in discussions of the arrow of time in quantum
cosmology \cite{Ze}.

What happens for other values of $\beta$? There exists a critical value,
$\beta_c=1/3$, for which the metric (4) is degenerate. For
$\beta<\beta_c$ it is positive definite, while for $\beta>\beta_c$
it is indefinite (the general relativistic case discussed above
is a special example). This becomes obvious if one introduces new
coordinates on $\mbox{Riem}\Sigma$ \cite{Ni},
\bea \tau &=& 4\sqrt{\left\vert\beta -\frac{1}{3}\right\vert}h^{1/4},
\nonumber\\
     r_{ab} &=& h^{-1/3}h_{ab}, \lb{6} \eea
since then the ``line element" in superspace can be written in the form
\be G_{\beta}^{abcd}dh_{ab}\otimes dh_{cd}=-\mbox{sgn}
    \left(\beta-\frac{1}{3}\right)d\tau\otimes d\tau
    +\frac{\tau^2}{16\vert\beta -\frac{1}{3}\vert}
    \mbox{Trace}(r^{-1}dr\otimes r^{-1}dr). \lb{7} \ee
One recognizes that the minus sign which occurs for $\beta>\beta_c$
is connected with the conformal part (``scale factor") of the
three-metric.

It is important to note that the theories with $\beta\neq 1$
do not correspond to a reparametrization invariant theory on the
Lagrangean level, if the three-metric is the only configuration
variable. The reason is that the validity of the commutation
relations of the generators of surface deformations
(which incorporate reparametrization invariance) necessarily
leads to general relativity, if the phase space only consists of the
three-metric and its momentum \cite{Ho}. In spite of this, it is
nevertheless possible that the theories for $\beta\neq1$
are physically viable. First, from the viewpoint of canonical quantum gravity,
it is clear that spacetime covariance is not a fundamental
concept. Second, reparametrization invariance can be restored
by enlarging the configuration space of the theory. This in fact happens,
for example, in the case of the Jordan-Brans-Dicke theory. In
situations where the kinetic term of the scalar field is negligible,
 one can restrict
oneself to the superspace part of the full configuration space.
The metric on this subspace is given by (3) with an effective
$\alpha=(\omega+1)/(2\omega+3)$ \cite{Ma}. This gives further support
to the study of such supermetrics.

In the following we will study the connection of the sign in the
Wheeler-DeWitt metric with the attractivity of gravity as well as possible
cosmological consequences. To this purpose we will discuss the sign
of the acceleration of the whole three-volume
$V\equiv\int\sqrt{h}d^3x$, which is a coordinate-independent
quantity. Moreover, since we use a Gaussian coordinate system
($N=1$ and $N^a=0$), test particles move on worldlines of constant
spatial coordinate, so that even the local quantity
$\sqrt{h}d^3x$ has a direct physical significance.

 We thus need the second time derivative
 of $\sqrt{h}$, which is found from the Hamilton equations of motion
\[ \dot{h}_{ab}=\{h_{ab}, H\} =32\pi G\ G_{abcd}\pi^{cd}, \]
and
\[ \dot{\pi}^{ab}=\{\pi^{ab}, H\}=-16\pi G
   \frac{\partial G_{nmpq}}{\partial h_{ab}}\pi^{nm}\pi^{pq}
   -\sqrt{h}M^{ab}-\frac{\sqrt{h}}{16\pi G}\left(G^{ab}
   +\Lambda h^{ab}\right), \]
where
\[ \sqrt{h}G^{ab}=-\frac{\delta}{\delta h_{ab}}
   \int\sqrt{h}Rd^3x=\sqrt{h}\left(R^{ab}-\frac{1}{2}h^{ab}R\right), \]
and
\[ \sqrt{h}M^{ab}=\frac{\delta}{\delta h_{ab}}\int\sqrt{h}
   {\cal H}_m d^3x= \frac{\partial\sqrt{h}{\cal H}_m}{\partial
   h_{ab}}. \]
In the last step we have assumed that the matter Hamiltonian
depends ultralocally on the metric.

After some calculations (which basically generalize the calculations
in section~6 of \cite{Wi}) we find
\[ \frac{d^2V}{dt^2}= (3\alpha-1) \int d^3x \left(-\frac{3}{8}
    G^{abcd}\dot{h}_{ab}\dot{h}_{cd}+\sqrt{h}h_{ab}
    (G^{ab}+\Lambda h^{ab}+16\pi GM^{ab})\right).\]
Using now the constraint equation
\[ G^{abcd}\dot{h}_{ab}\dot{h}_{cd}= 4\sqrt{h}(R-2\Lambda
   -16\pi G{\cal H}_m), \]
we arrive at our final result
\be \frac{d^2V}{dt^2}= -3(3\alpha -1)\int d^3x\sqrt{h}\left(\frac{2}{3}
    R-2\Lambda-16\pi G({\cal H}_m-\frac{1}{3}h^{ab}\frac{\partial
    {\cal H}_m}{\partial h^{ab}})\right). \lb{8} \ee
Gravity is attractive if the sign in front of the integral is negative.
This can be recognized from an inspection of the various terms
in (8):
Gravity is attractive if a positive Ricci scalar contributes with
a negative sign to the acceleration (8). This is the case for $\alpha
>\alpha_c\equiv 1/3$, which is the critical value for the
 metric (3). A cosmological constant $\Lambda >0$ then acts
 repulsively. As far as the coupling to the matter terms in (8) is
 concerned, an overall change of sign corresponds to
 a change of sign in the gravitational constant, which would become
 negative for $\alpha<\alpha_c$. The fact that the matter terms
then contribute with a positive sign to the acceleration of
the volume should not be disturbing, since the local scale
itself (the third root of the volume) is decelerating. The important
fact is the overall change of sign in (8) for $\alpha<\alpha_c$
compared to general relativity.

  Thus, {\em the attractivity of gravity is
intimately connected with the indefinite signature of the
Wheeler-DeWitt metric}.
The contribution of the matter terms will now be discussed in more detail
in the context of specific examples. Consider, e.g., the Hamiltonian
density for a scalar field,
\be {\cal H}_m =\frac{\pi^2_{\phi}}{2h} +\frac{1}{2}
    h^{ab}\phi_{,a}\phi_{,b}+\frac{1}{2}m^2\phi^2+V(\phi). \lb{9} \ee
{}From (8) one finds in this case
\be \frac{d^2V}{dt^2}=-2(3\alpha-1)\int d^3x\sqrt{h}
    \left(R-3\Lambda-24\pi G(\frac{1}{2}m^2\phi^2+ \frac{1}{3}
    h^{ab}\phi_{,a}\phi_{,b}+
    V(\phi))\right).
    \lb{10} \ee

Since the matter terms here contribute with
a positive sign in (8)
for $\alpha>\alpha_c$, only a positive curvature term can lead
to a deceleration of the volume. The matter terms
can contribute, however, with a minus sign in the case of
higher order tensor fields, since then the ``pressure-like"
derivative term of ${\cal H}_m$ with respect to the metric may
dominate.\footnote{
We note that for phenomenological matter, for which the kinetic
term is neglected,
 ${\cal H}_m -1/3
h^{ab}\partial{\cal H}_m/\partial h^{ab}
\equiv (\rho-p)/2$, where $\rho$ and $p$ are density and pressure,
respectively.}

We conclude with a discussion of possible cosmological consequences.
 For general values of $\alpha$ the Friedmann equation for
the scale factor $a$ reads
\be \frac{\dot{a}^2}{2(3\alpha-1)} =-k+
    \frac{8\pi G}{3}a^2\left(\rho+\frac{\Lambda}{8\pi G}\right).
    \lb{11} \ee
We recognize that modifying $\alpha$ modifies the expansion rate
of the Friedmann universe. In
the case of a radiation dominated universe (11) reads
\be \frac{\dot{a}^2}{2a^2(3\alpha-1)}\approx \frac{8\pi G}{3}\rho. \lb{12} \ee
Thus, the influence of $\alpha$ can be described by an
 effective density (taking into account photons, neutrinos, and
 electrons)\footnote{We assume that the temperature is much
 bigger than the electron mass.}
\[ \rho_{eff}=2(3\alpha-1)a_B(1+\frac{7}{8}N_{\nu}
   +\frac{7}{8}\cdot 2)T^4, \]
where $N_{\nu}$ is the number of neutrinos, $T$ the temperature,
and $a_B$ is the Stefan-Boltzmann constant. Since one knows from LEP
that the number of massless neutrinos is three, the observed
amount of primordial helium in the universe, $Y=0.22\pm0.03$,
gives restrictions on the actual value of $\alpha$: If $\alpha$
is too big (too small), too much (too little) helium is produced.
We can estimate that for the above value of $Y$ (see, e.g., \cite{Pe})
$\alpha$ must, roughly, lie between $0.4$ and $0.55$.

\vskip 5mm

We are grateful to Alvaro Llorente for correspondence and critical
remarks.

\end{document}